\documentclass{elsart}

 \usepackage{graphics}
\usepackage{amssymb}

\begin{document}

\begin{frontmatter}


\begin{center}
\title{Mixed state Hall effect in the multiphase superconductors}


\author{I.\ Jane\v cek$^1$, P. Va\v sek$^2$}

\address
{$^1$University of Ostrava, Dvorakova 7, 70103 Ostrava, Czech Republic\\
$^2$Institute of Physics ASCR, Cukrovarnick\'a 10,\\ 162 53 Praha
6, Czech Republic\\}

\end{center}
\begin{abstract}
% Text of abstract
Hall effect below T${_c}$ in multiphase superconductors has been studied on Bi-based superconductors.
Samples with different relative content of 2212 and 2223 phase have been prepared.  The phase content has been verified by X-ray diffraction. Results show that while resistance and ac susceptibility is almost insensitive to the content of 2212 phase, the qualitative behavior of the Hall resistance is strongly influenced by the presence of both phases. Theoretical calculation of Hall resistance has been made based on effective medium approximation and compared with experimental results.

\end{abstract}

\begin{keyword}
superconductors, Hall effect

\PACS 72.15.Gd \sep 74.72.-h \sep 74.25.Fy \sep 74.60.Ec 
\end{keyword}
\end{frontmatter}

% main text
\section{Introduction}
Preparation of  single phase  high T${_c}$ superconductors is in many cases difficult technological task, in particular in those with the highest critical temperatures (Bi-, Tl- and Hg-based materials). In such superconductors presence of minority phase can influence experimental results and afterwards the interpretation of them as well.
Theoretical study of the Hall effect in the mixed state of high T${_c}$
superconductors did not bring any fully satisfactory explanation of observed experimental results.
 From the time of classical work of Bardeen et al \cite{Ba} and Nozieres et al \cite{No} a lot of theoretical work has been published relating the Hall voltage behavior to the flux pinning
\cite{A} , backflow of thermally excited quasiparticles \cite{B}, layered structure of superconductors \cite{c}, a vortex glass transition \cite{D}, imbalance of the electron density between center and outer region of the vortex \cite{E} or to the interaction among normal state charge carriers, superconducting fluid and vortices \cite{F}. A main feature of the Hall voltage is the change of
sign with the temperature. One change of sign with decreasing temperature from above T${_c}$
has been observed in YBaCuO superconductors \cite{G}, while in Bi- and Tl- based ones two
 changes of sign have been observed \cite{H}. Recently even tripple change has been observed on Hg samples \cite{I}. With increasing magnetic fields the Hall voltage generally moves to  positive values, the change of sign disappears but the number of local minima does not change.
In this paper we have studied how the mixed state Hall voltage will change if  other minority superconducting phase will be present in the sample under study.\\
This problem has been studied theoretically on the basis of the effective medium theory and experimentally through resistivity and Hall effect data measured on chosen bi-phase superconducting materials.\\   
As a material under study BiSrCaCuO system has been used. The reason for this is that it is rather difficult to prepare a single phase sample and therefore almost always the presence of 2223 and 2212 phases can be present which can influence the  results obtained and therefore also the theoretical interpretation. 

\section{Theory}

Only few papers exist describing transport properties of binary mixture i.e. situation where small pieces of a phase 1 are mixed with similar pieces of the second phase 2.  Very useful model for such situation is the effective-medium approach originally developed by Bruggeman \cite{Brug}. This method is designed to deal with the inhomogeneous media in which different phases are randomly distributed in the form of grains of an arbitrary shape, size and orientation. But the formalism is greatly simplified if an assembly of ellipsoids is supposed. Using this model one can calculate effective conductivity for uncorrelated (no correlation between the positions of the grains of different phase) binary mixture \cite{Lan}. For spherical inclusions one can get
\begin{eqnarray}
\hspace*{-.8cm}\sigma{_{eff}} = \frac{(3x{_1}-1)\sigma{_1}+(3x{_2}-1)\sigma{_2}+\sqrt{((3x{_1}-1)\sigma{_1}+(3x{_2}-1)\sigma{_2})^2+8\sigma{_1}\sigma{_2}}}{4}
\end{eqnarray}
where $\sigma{_1}$,$x{_1}$ and $\sigma{_2}$,$x{_2}$ are conductivities and fractional volumes of the first and the second component, respectively.
The effective Hall mobility was calculated by  Witt in \cite{wit} where he obtained 
\begin{equation}
\mu^H{_{eff}}= \frac{A{_1}\mu^H{_1}x{_1} -A{_2}\mu^H{_2}x{_2}}{1-2(A{_1}x{_1}+A{_2}x{_2})}
\end{equation}
with $A{_k}=(3\sigma{_{eff}}\sigma{_k})/(2\sigma{_{eff}}+\sigma{_k})^2$,  k=1,2.

$\mu^H{_1}$ and $\mu^H{_2}$ are Hall mobilities of both subsystems.\\
 By the application of these formulas the ohmic and Hall part of the effective resistivity tensor can be derived.
\begin{equation}
\rho=\frac{4}{(X{_1}/\rho{_1}+X{_2}/\rho{_2}+\sqrt{(X{_1}/\rho{_1}+X{_2}/
\rho{_2})^2+8/\rho{_1}\rho{_2}}}
\end{equation}
where $X{_k}=3x{_k}-1$ for k= 1,2.
\begin{eqnarray}
\rho{^H}= \frac{\rho}{1-2(Y{_1}+Y{_2})}(\frac{Y{_1}}{\rho{_1}}\rho^H{_1}+\frac{Y{_2}}{\rho{_2}}\rho^H{_2})
\end{eqnarray}
where $Y{_k}= 3\rho\rho{_k}x{_k}/(2\rho + \rho{_k})$
and $\rho^H{_k}$ for k=1,2 represent Hall resistivities of both phases, respectively. 
The well known result about percolation threshold for the effective resistivity of superconductor ($\rho{_2}=0$) with admixture of nonsuperconducting particles ($\rho{_1}\neq 0$) follows from Eq. 3. The resistivity reaches zero value if the superconducting path is established through the sample. This situation takes place when 1/3 of total volume is occupied by superconducting particles (valid for spherical grains) (see Fig. 1). For general ellipsoids  another  volume fraction is needed to established superconducting percolation path.\\     
The analysis of the formula for the Hall  resistivity is not so straightforward. While in the case of resistivity only two parameters are present ($\rho{_k}$ for k=1,2), formula for $\rho{^{H}}$ includes moreover Hall resistivities for both components. 
At first, we consider case of sample with uniform Hall resistivity but with different resistivities of both component to get an image of influence of these resistivities on effective Hall resistivity (Fig. 2). We find that the effective Hall resistivity of the mixture is lower than  values of Hall resistivities for both components. The effective Hall resistivity has minimum for percolation threshold. However, value of this minimum is positive for arbitrary ratio $\rho{_2}/\rho{_1}$. Lowest value approaches  $1/4$ for $\rho{_2}/\rho{_1}\rightarrow 0$. 
The sign of the effective Hall resistivity is  determined by the  second term in Eq. 4 because the first term , which depends only on resistivities, can not reach  negative values. In Fig. 2 lines of effective Hall resistivity as a function of volume fraction are drawn for several values of the  ratio of $\rho{_2}$ to $\rho{_1}$ supposing positive $\rho^H{_1}$ = $\rho^H{_2}$. The change of sign  is illustrated in the  Fig. 3 where the same is made for different ratio of $\rho^H{_2}/ \rho^H{_1}$ at $\rho{_2}/ \rho{_1}\rightarrow 0$. As was mentioned above for the effective resistivity of mixture of superconducting and nonsuperconducting particles percolation threshold exists at volume fraction of superconducting phase $x{_2}=1/3$. It seems from   Fig. 3  that there is no such behavior for effective Hall resistivity with $\rho{_2}/\rho{_1} \rightarrow 0$. However it holds for any nonzero ratio  $\rho^H{_1}/\rho^H{_2}$. But one should have in mind that superconductivity with zero resistivity is also supplemented by zero Hall resistivity and so the percolation threshold will be also observed in the Hall resistivity data.\\ 
Until now we have studied only concentration dependence of effective resistivities supposing that resistivities of both components do not change. In the next we would like to use the model of effective mediums to investigate the temperature dependence of the effective resistivities. In this case we will assume constant $x{_2}$, which is determined by chemical composition and we will use temperature dependence of pure single component resistivities to determine the  effective ones. 
Calculated temperature dependencies  of the effective Hall resistivity are presented on Fig.4.
The calculation is based on the hypothetical temperature dependence (but the qualitative features are in accordance with experiment) of  resistivities 
and Hall resistivities describing the behavior of the single 2223 and 2212 phases of Bi-based superconductors. For mixtures of phases  curves with double minima are obtain.

\section{Experimental results and discussion}

For the experimental study Bi(Pb)SrCaCuO polycrystalline samples with varying volume fraction of
low $T{_c}$ phase (2212) and high $T{_c}$ phase (2223) has been prepared
 together with  pure single 2223 phase sample by special thermomechanical treatment \cite{plech}.  The ratio of phases has been controlled by different annealing  time at 840 $^o$ C and 830 $^o$ C. Shorter time of annealing means lower content of 2223 phase. 
Relative volumes of superconducting phases have been determined by X-ray diffraction studies. In addition to the resistivity and Hall effect  the ac and dc susceptibility has been also determined.\\
Resistances and Hall resistances in magnetic field 4 T for the samples with 2223 phase volume fraction 100\%, 75\%, 50\%  are shown in Fig. 5. One can see that while single phase samples reveal only  one local minima in the temperature dependence of the Hall resistance,  samples in which both phases are present show two local minima before going to zero at low temperatures. Comparison with Fig. 5 reveals qualitative agreement with calculated curves based on the effective medium model.
However one can  see that the experimental  curves quantitatively disagree with presented theoretical model. The reason for this is probably connected with the roughness of the model used. 
The roughness of the mentioned model follows at least from the assumption of   spherical shape of grains, which is certainly not the case of high $T{_c}$ superconductors. In our calculation we also suppose that different phases are not correlated which can not be  fulfilled as well.  Moreover it is well known \cite{Si} that the volume fractions of  different phases determined from the resistivity and susceptibility  measured at very low magnetic fields do not agree with those determined by X-ray diffraction. While the two above mentioned measurements can not "see" lower phase if its grains are covered by a thin layer of 2223 phase, the X-ray can.

\section{Conclusion}

The formulas for the effective Hall resistivity have been derived on the basis of the effective medium approach. The calculated temperature dependence has been compared with the experimentally obtained curves and qualitative agreement has been obtained. The results confirmed that in the multiphase samples several local minima can be observed in the temperature dependence of the Hall resistivities due to the presence of twophases with different  Hall resistivities. This means, that not only the mechanismus described by Kopnin \cite{Ko} can be responsible for observed results in \cite{I}. From this work follows that one must be careful in the interpretation of the qualitative features of experimentally determined Hall effect. However,  one must have in mind that the used model is too simple to get quantitative agreement.

\section{Acknowledgements}

This work has been supported by GACR under project No. 202/00/1602 and by GAAS under project No. A1010919/99  and by Ministry of Education  under research plan No. 173100003   
%\begin{figure}[t]
%\scalebox{.25}{\includegraphics[-350,110][300,400]{Fig1.eps}}
%\vspace*{3cm}
%\caption{pokus o text k obrazku}
%\end{figure}

\label{}

\newpage
{\large Figure captions}

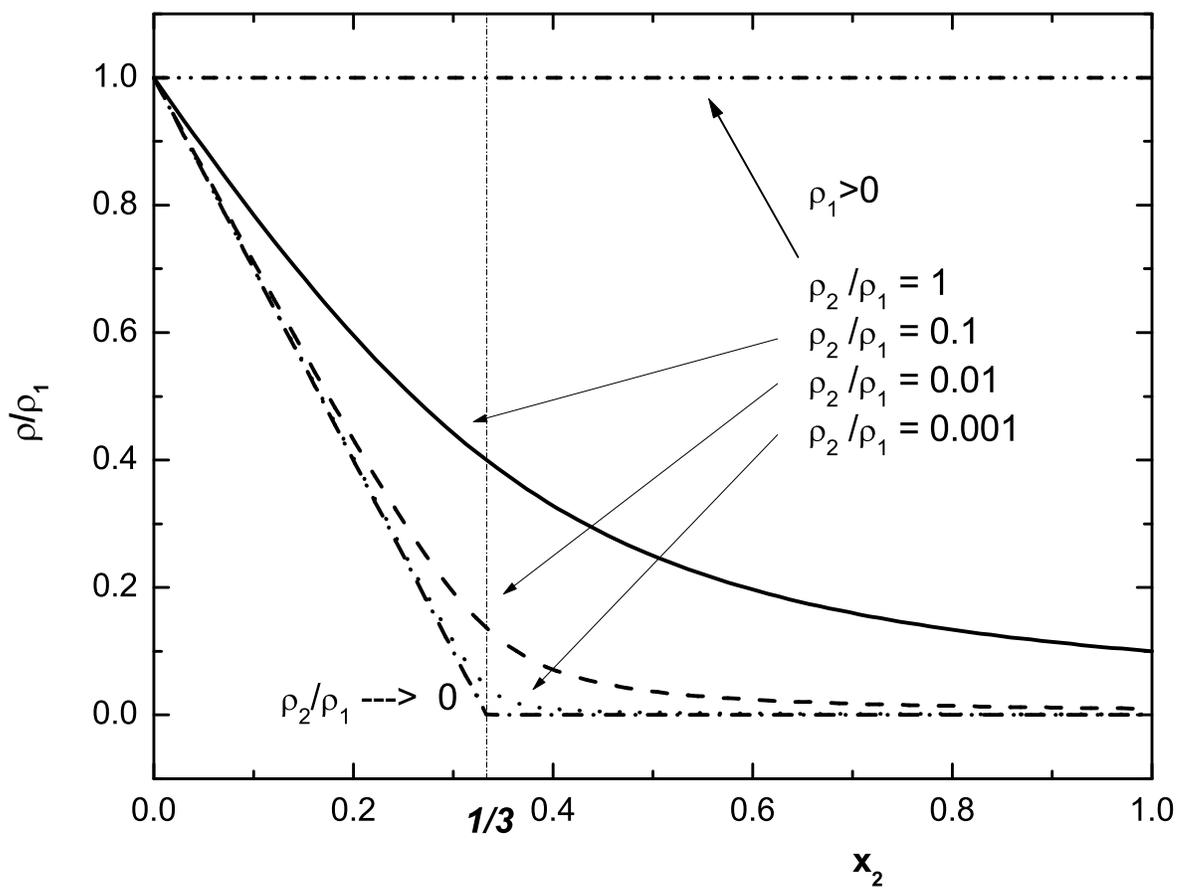
Fig. 1: Normalized effective resistivity $\rho/\rho{_1} $
as a function of volume fraction 
$x{_2}$ for different ratio of single phase resistivities $\rho{_2}/\rho{_1}$ .\\

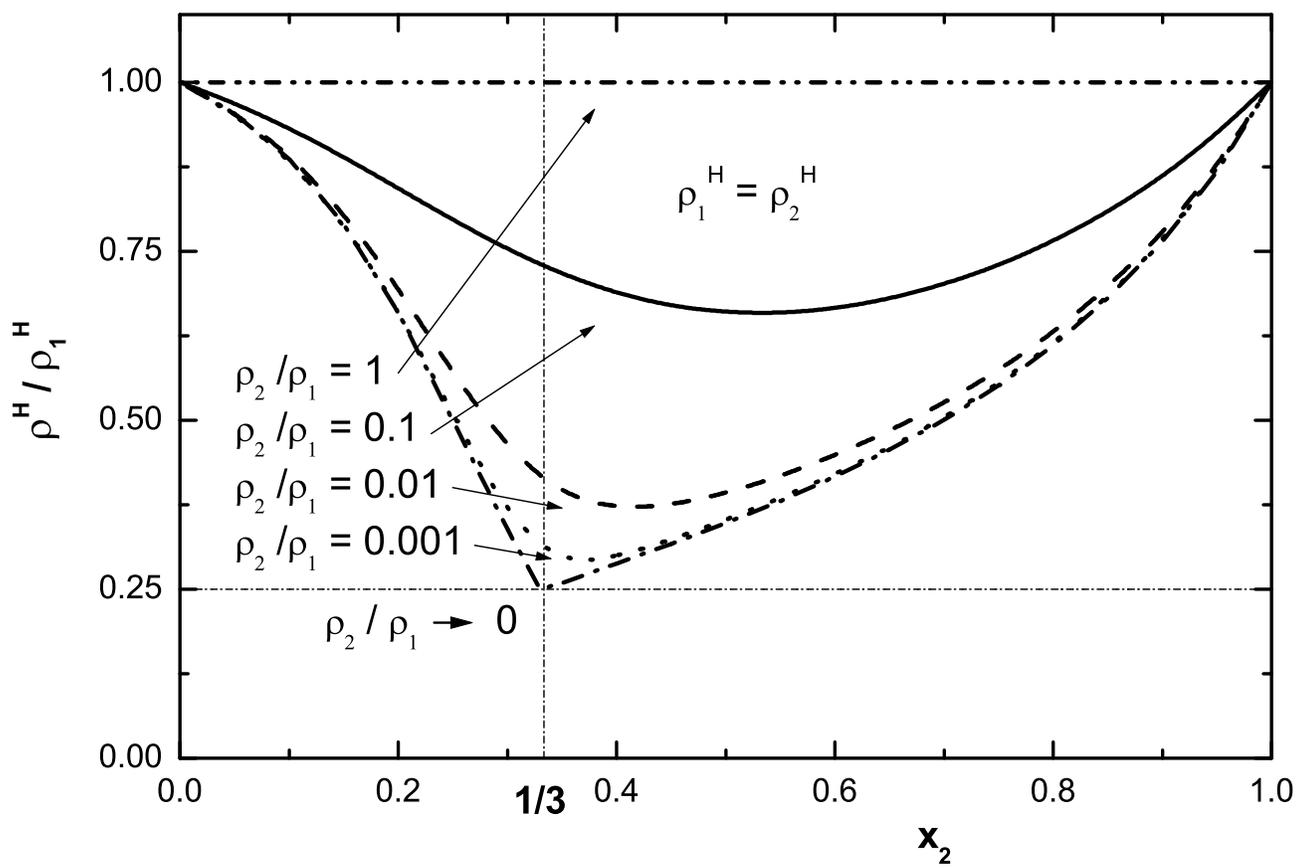
Fig. 2: Normalized effective Hall resistivity $\rho^H/\rho{_1}^H $
as a function of volume fraction $x{_2}$ for different ratio of single phase resistivities $\rho{_2}/\rho{_1}$ and equal Hall resistivities ($\rho{_2}^H/\rho{_1}^H$= 1).\\

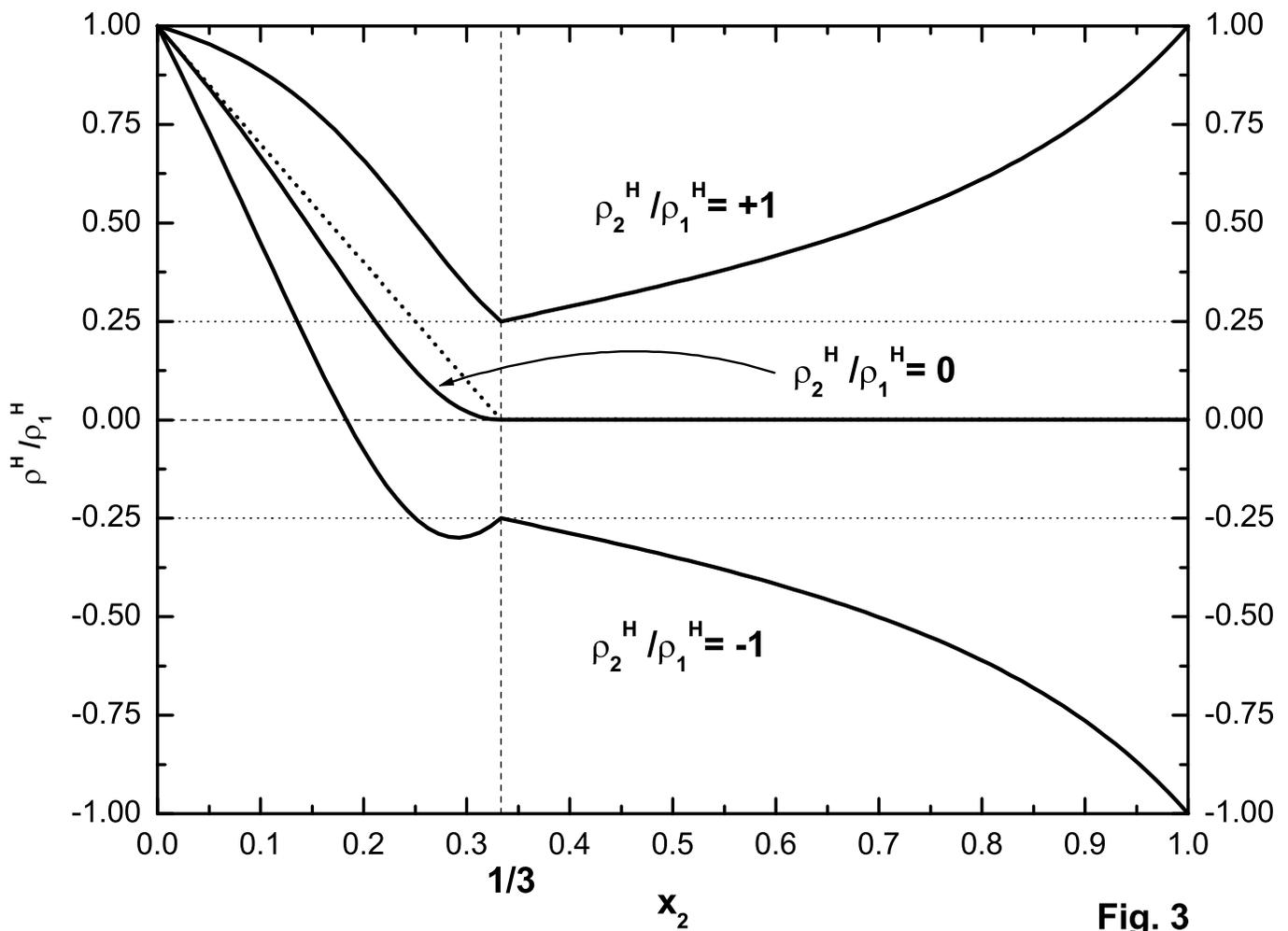
Fig. 3: Normalized effective Hall resistivity $\rho^H/\rho{_1}^H $
as a function of volume fraction $x{_2}$ for parameter $\rho{_2}/\rho{_1} \rightarrow 0$ and parameter $\rho{_2}^H/\rho{_1}^H$  equal +1,0,-1, respectively. For comparison the dot  curve for  $\rho/\rho{_1} $ is shown .\\

Fig. 4: Calculated temperature dependencies of effective resistivity (a) and effective Hall resistivity (b) for different volume fraction  based on assumed temperature dependence of single phase samples (0\%,100 \%). The presented curves correspond to low  magnetic field where sign changes of the Hall resistivity are usually observed.\\

Fig. 5:  Temperature dependencies of normalized resistivity (a) and Hall resistivity (b) for samples with different volume fraction of 2223 phase measured at 4T ( in (a) : $\bigcirc$ - 100 \%,  $\Box$ -75\%,$\triangle$ - 50 \%).\\

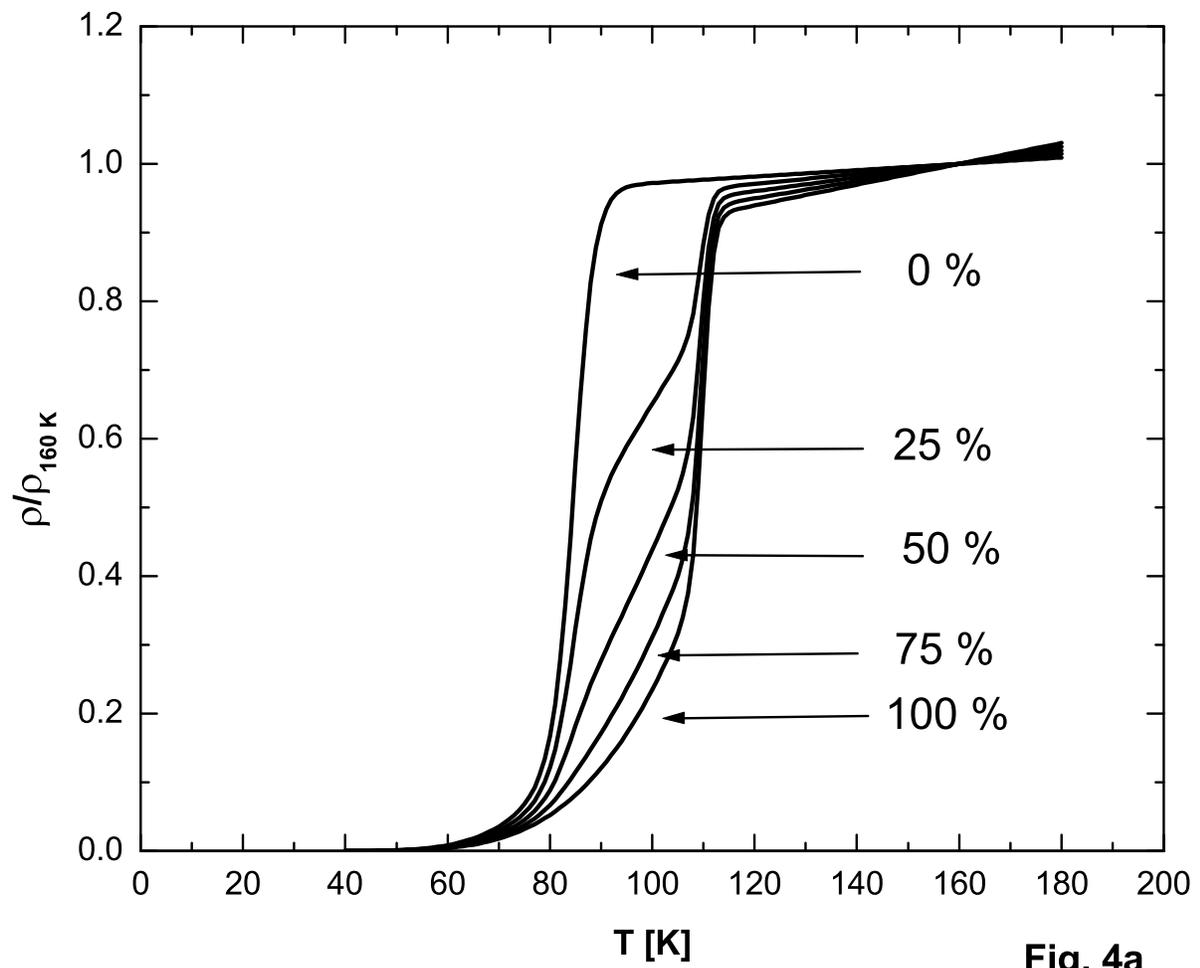
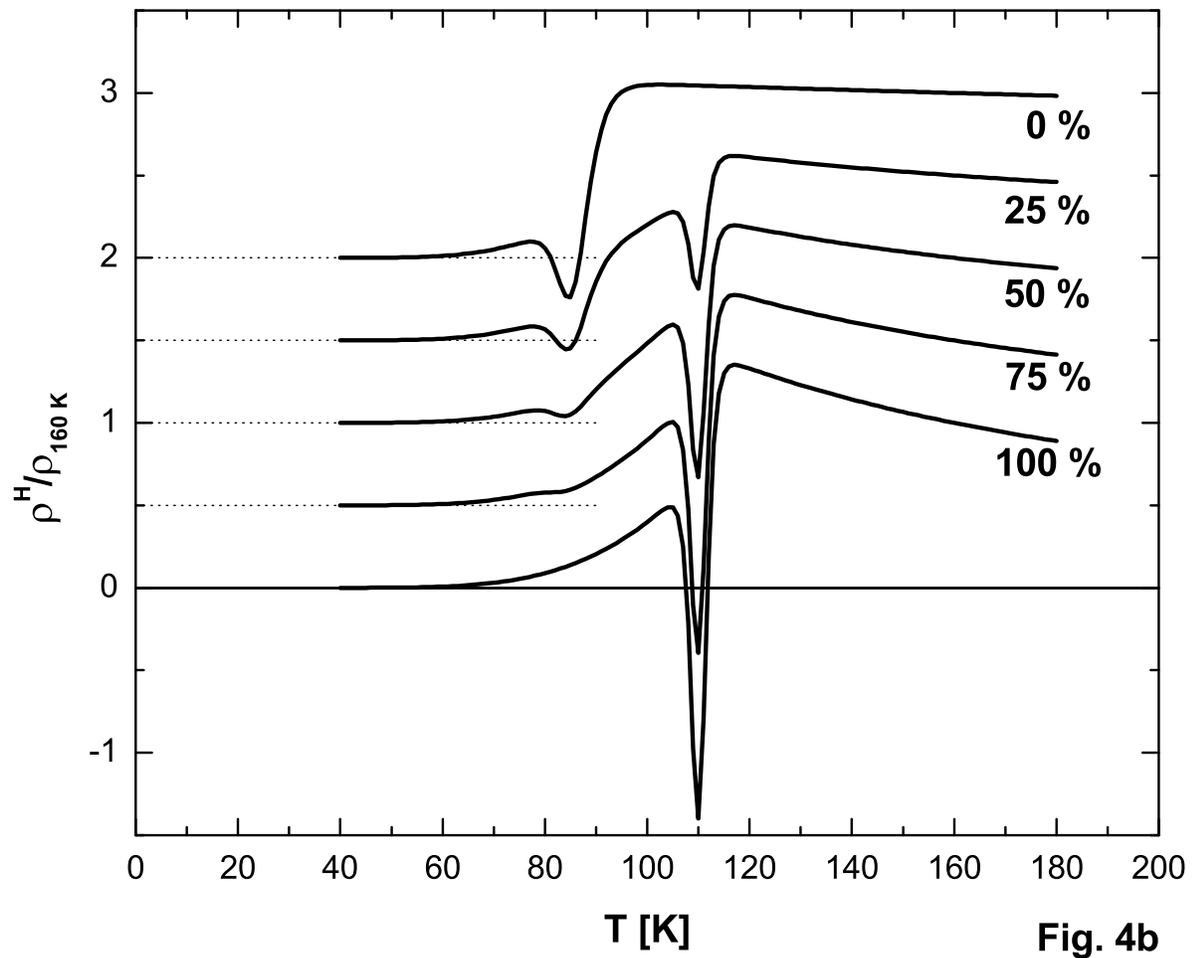
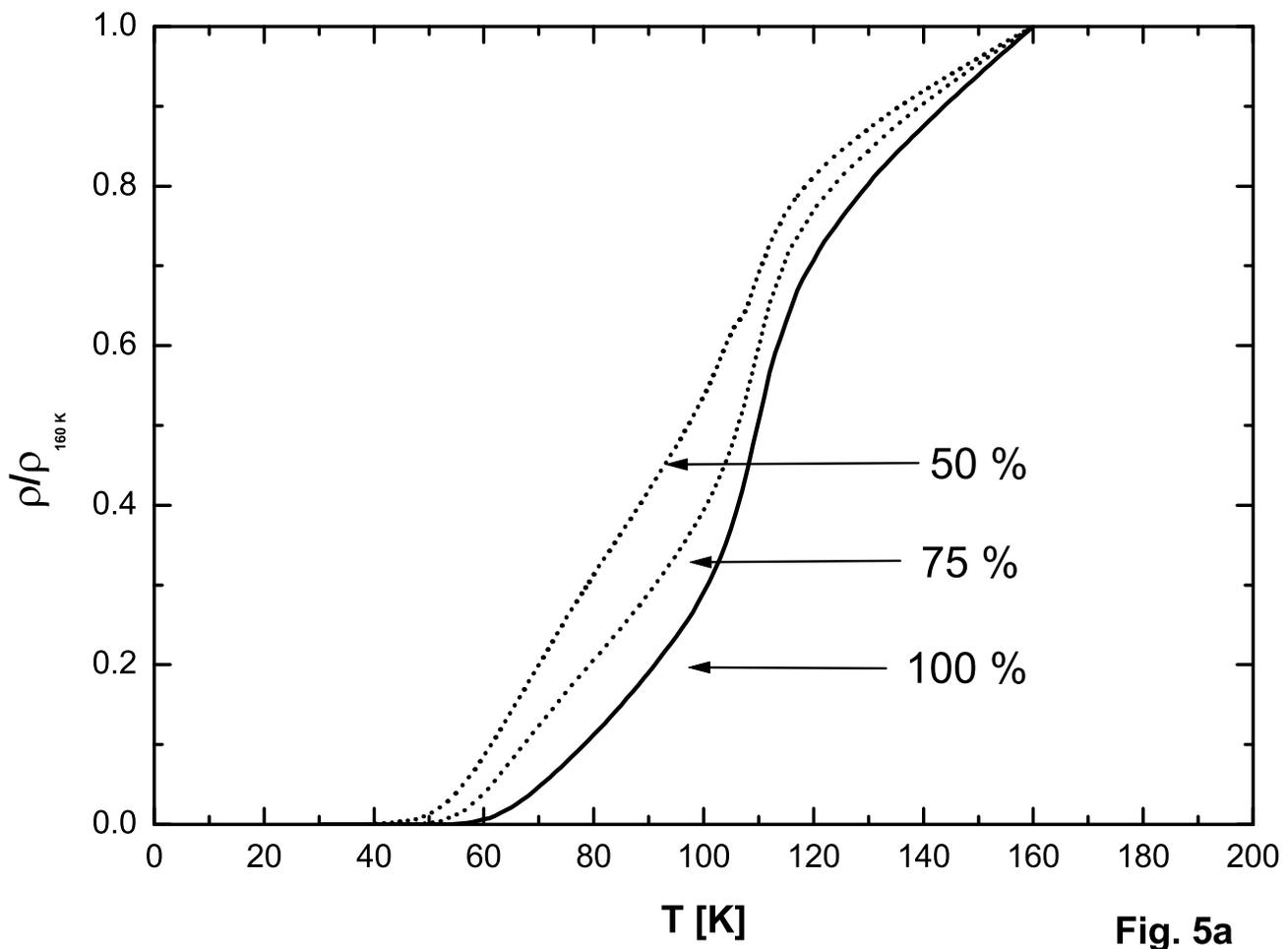
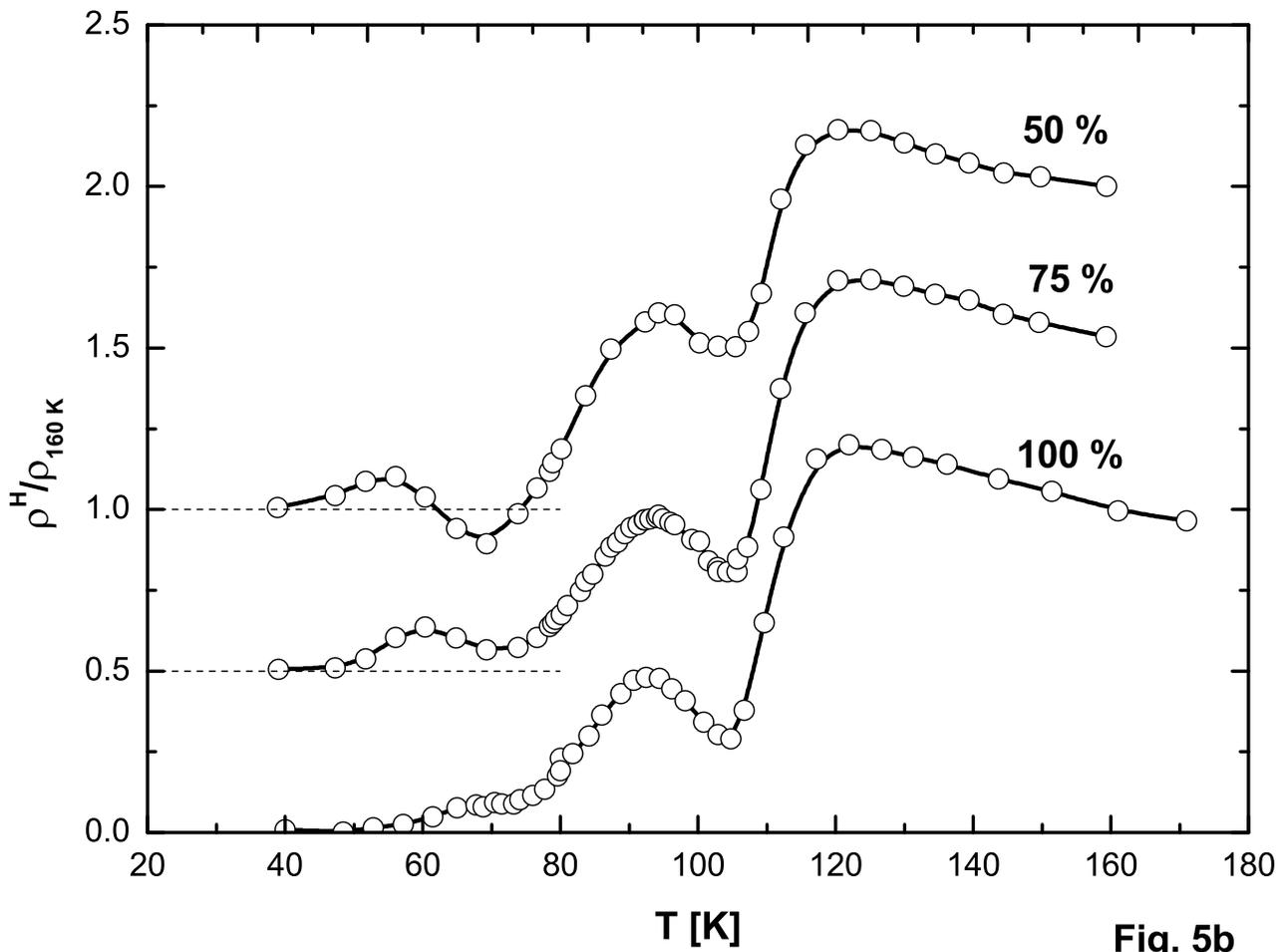
\end{document}